\documentclass[letterpaper, 10 pt, conference]{ieeeconf}  % Comment this line out if you need a4paper

\IEEEoverridecommandlockouts                              % This command is only needed if 
\usepackage{amsmath,amssymb,mathtools}
\usepackage{graphicx} 
\usepackage{xspace}
\usepackage{pifont}% http://ctan.org/pkg/pifont
\usepackage{wasysym}
\usepackage{url}
\usepackage[dvipsnames]{xcolor}
\usepackage{makecell}
\usepackage{multirow}
\usepackage{wrapfig}
\usepackage{nicematrix}
\usepackage{tikz}
\usepackage{adjustbox}
\usetikzlibrary{shapes,arrows,calc,positioning,graphs,quotes,automata}
\usepackage{pgf}
\usepackage{pgfplots}
\pgfplotsset{compat=1.18}
\usepackage{tcolorbox}
\usepackage{booktabs,tabularx,makecell}
\usepackage{cuted,caption,booktabs,tabularx,array}
\usepackage{tabularx, makecell, booktabs}
\newcolumntype{C}{>{\centering\arraybackslash}X}
\newcolumntype{L}{>{\raggedright\arraybackslash}X}

\let\labelindent\relax
\usepackage[shortlabels]{enumitem}

\usepackage{stmaryrd} 
\usepackage{cuted}     % for \begin{strip}...\end{strip}
\usepackage{caption}   % for \captionof outside floats
\usepackage{booktabs}  % for \toprule \midrule \bottomrule
\usepackage{tabularx}  % for tabularx and X columns
\usepackage{array}     % for \newcolumntype and \arraybackslash
\usepackage{amsmath,amssymb} % robust math
\usepackage{mathtools} % (optional) nicer math
\usepackage{cleveref}
\usepackage{bm}
\usepackage{algorithm}
\usepackage{tikz}
\usetikzlibrary{arrows.meta,positioning,bending}
\usepackage{subcaption}
\usepackage{algorithmic}
\usepackage{dblfloatfix}

\newcolumntype{Y}{>{\raggedright\arraybackslash}X}

\newcommand{\jay}{\mathrm{J}}    % any intersection

\usepackage{xcolor}
\usepackage{booktabs}
\usepackage{array}      
\usepackage{cuted}    % in-place full-width content in two-column docs
\usepackage{capt-of}  % \captionof{table}
\usepackage{tabularx,booktabs,makecell,array}
\usepackage{subcaption}
\usepackage{algorithm}

\usepackage{booktabs}
\usepackage{multirow}
\usepackage{adjustbox}
\usepackage{makecell}
\usepackage{placeins}
\usepackage[table]{xcolor}
\definecolor{lightgray}{gray}{0.95}

\tikzset{
   operator/.style = {rectangle, thick, draw, rounded corners, minimum width=0.6cm, minimum height = 1cm},
   choice/.style = {diamond, thick, draw, inner sep=0},
   sgvertex/.style = {draw, ellipse, minimum width=3cm, minimum height=1cm, thick},
}

\newcolumntype{C}{>{\centering\arraybackslash}X}

\newcommand{\carlabase}{\ensuremath{\mathbf{CARLA_{\mathrm{base}}}}\xspace}
\newcommand{\carlaTenX}{\ensuremath{\mathbf{CARLA_{10\times}}}\xspace}

\IEEEoverridecommandlockouts 
\newcommand{\ltlf}{\ensuremath{\mathrm{LTL}_{\!f}}\xspace}

\newcommand{\stada}{\textsc{STADA}\xspace}

\newcommand{\sceneflow}{\textsc{SceneFlow}\xspace}
\newcommand{\scenic}{\textsc{Scenic}\xspace}

\newcommand{\ignore}[1]{}

\title{\LARGE \bf
\stada: Specification-based Testing for Autonomous Driving Agents
}

\begin{document}

% \keywords{runtime monitoring, runtime enforcement, scene graph, autonomous vehicles}

\author{Joy Saha$^{1}$, Trey Woodlief$^{2}$, Sebastian Elbaum$^{1}$, and Matthew B. Dwyer$^{1}$
\thanks{$^{1}$University of Virginia, Virginia, USA \{daa7mv, selbaum, matthewbdwyer\}@virginia.edu$\qquad^{2}$William \& Mary, Virginia, USA woodlief@wm.edu}%
}

\maketitle
\begin{abstract}
Simulation-based testing has become a standard approach to validating autonomous driving agents prior to real-world deployment. 
A high-quality validation campaign will exercise an agent in diverse contexts comprised of varying static environments, e.g., lanes, intersections, signage, and dynamic elements, e.g., vehicles and pedestrians.
To achieve this, existing test generation techniques rely on template-based, manually constructed, or random scenario generation.
When applied to validate formally specified safety requirements, such methods either require significant human effort or run the risk of missing important behavior related to the requirement.

To address this gap, we present \stada, a \underline{S}pecification-based \underline{T}est generation framework for \underline{A}utonomous \underline{D}riving \underline{A}gents that systematically generates the space of scenarios defined by a formal specification expressed in 
% linear
temporal logic (\ltlf). 
Given a specification, 
\stada constructs all distinct initial scenes, a diverse space of continuations of those scenes, and
% code to instantiate simulation scenarios
simulations that
reflect the behaviors of the specification.

Evaluation of \stada on a variety of \ltlf specifications formalized in \sceneflow using three complementary coverage criteria demonstrates that \stada yields more than 2$\times$ higher coverage than the best baseline on the finest criteria and a 75\% increase for the coarsest criteria.  Moreover, it matches the coverage of the best baseline with 6 times fewer simulations.
While set in the context of autonomous driving, the approach
is applicable to other domains with rich simulation environments.

% The performance of \stada shows its potential utility in edge case bug detection.
\end{abstract}

\begin{figure*}[th!]
  \centering

\begin{subfigure}[b]{0.198\textwidth}
  \centering
  \resizebox{\linewidth}{3.3cm}{%
    \begin{tikzpicture}[
      >=Stealth,
      n/.style={
        circle,
        draw=black,
        fill=white,
        line width=0.8pt,
        minimum size=12mm,
        inner sep=0pt,
        font=\small
      },
      e/.style={->, draw=black, line width=0.9pt},
      lbl/.style={font=\scriptsize, fill=white, inner sep=1.2pt}
    ]

      % Fixed bounding box (ensures full usage of subfigure area)
      \path[use as bounding box] (-3,-2.5) rectangle (3,3);

      % Nodes (triangle layout)
      \node[n] (ego)  at (0,2.2) {ego};
      \node[n] (lane) at (-2,-1.5) {Lane1};
      \node[n] (bike) at (2,-1.5) {Bike1};

      % Edges (straight, symmetric)
      \draw[e] (ego) -- (lane)
        node[lbl, midway, left] {inLane@I \& inLane@F};

      \draw[e] (bike) -- (lane)
        node[lbl, midway, below] {inLane@I \& inLane@F};

    % Two separate ego → bike edges (visually separated)
    
    % ego -> bike (upper arc)
    \draw[e, bend left=25] (ego) to
      node[lbl, midway, above] {!near\_col@I \& behind@I}
      (bike);
    
    % bike -> ego (lower arc)
    \draw[e, bend left=25] (bike) to
      node[lbl, midway, below] {!toDRightOf@I \& behind@F}
      (ego);

    \end{tikzpicture}
  }
  \caption{Relational Graph (RG)}
  \label{fig:rg}
\end{subfigure}\hfill
  \begin{subfigure}[b]{0.198\textwidth}
    \centering
    \includegraphics[width=\linewidth,height=3.3cm]{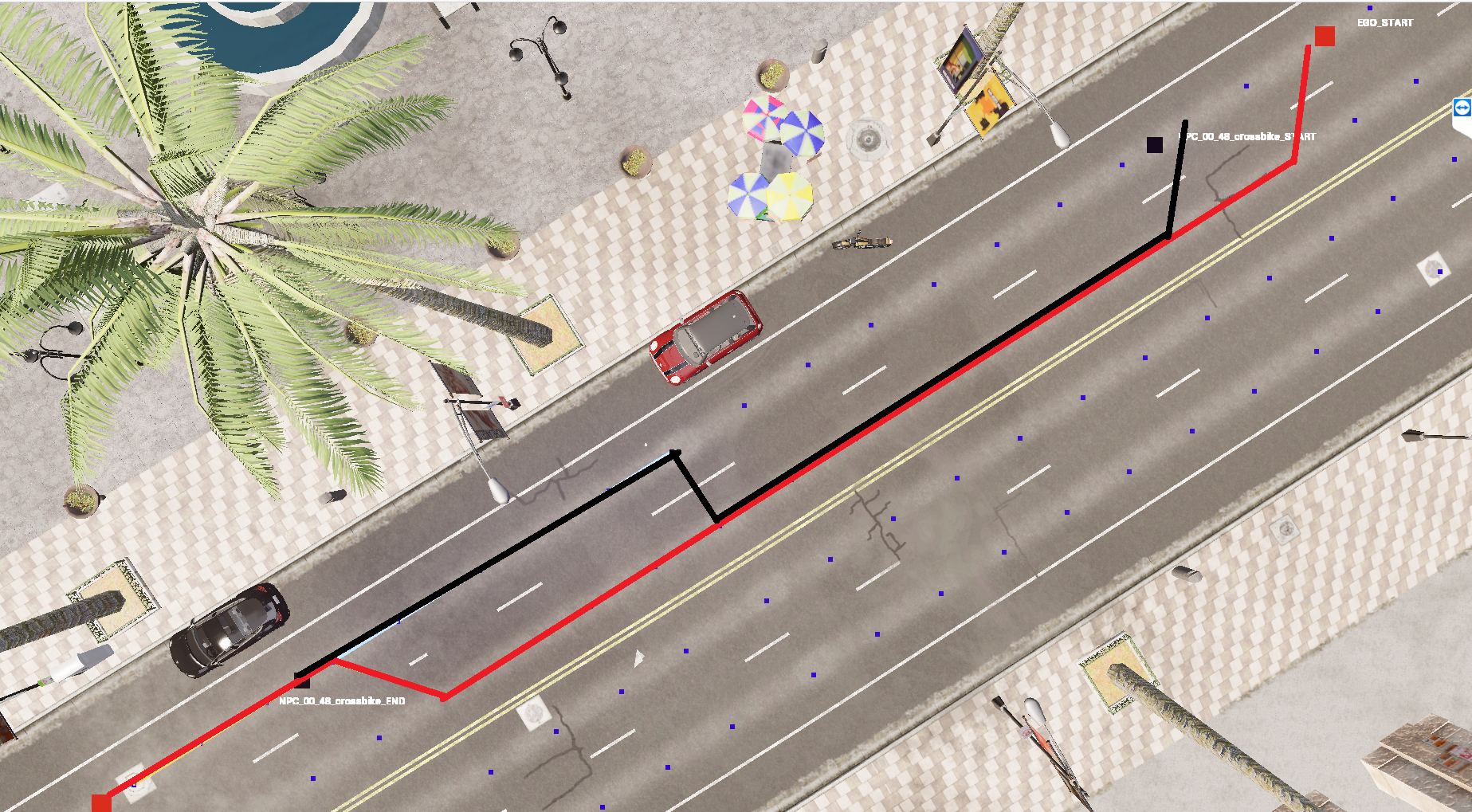}
    \caption{Ego and NPC paths}
    \label{fig:init_back}
  \end{subfigure}\hfill
  \begin{subfigure}[b]{0.198\textwidth}
    \centering
    \includegraphics[width=\linewidth,height=3.3cm]{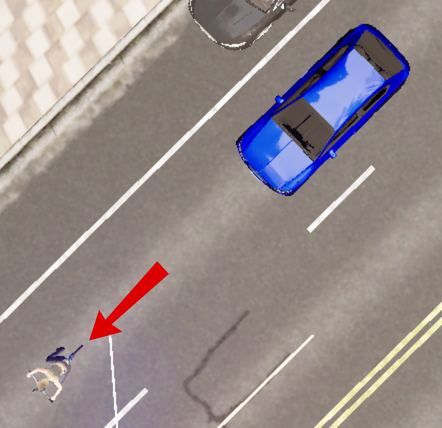}
    \caption{Initial Scene}
    \label{fig:init_top}
  \end{subfigure}\hfill
  \begin{subfigure}[b]{0.198\textwidth}
    \centering
    \includegraphics[width=\linewidth,height=3.3cm]{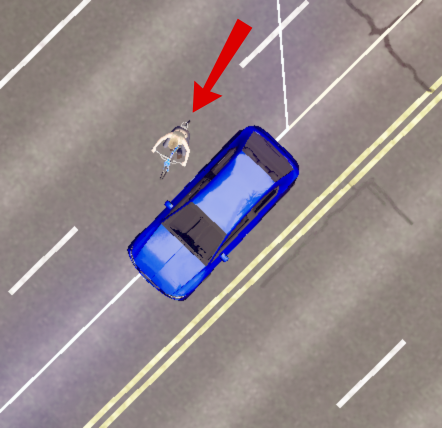}
    \caption{Passing Maneuver}
    \label{fig:passing}
  \end{subfigure}\hfill
  \begin{subfigure}[b]{0.198\textwidth}
    \centering
    \includegraphics[width=\linewidth,height=3.3cm]{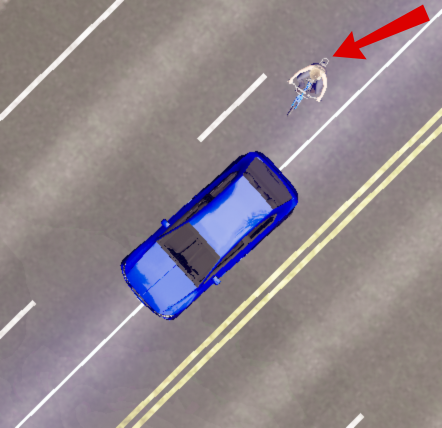}
    \caption{After Passing}
    \label{fig:passed}
  \end{subfigure}

  \vspace{2pt}
{\footnotesize
\begin{minipage}{0.98\textwidth}
\centering
% $\varphi \;=\;
% \underbrace{behind(ego,\, bike)}_{\textsf{A}}
% \wedge
% \underbrace{safeDistance_D(ego,\, bike)}_{\textsf{B}}
% \wedge
% \underbrace{\mathcal{F}(behind(bike,\, ego))}_{\textsf{C}}.$

% Full formula
\[
\varphi \equiv
\textcolor{blue}{
(
behind\_bike
\wedge
bike\_safe\_distance
\wedge
\mathbf{F}(in\_front\_bike) 
)
}
\Rightarrow
\mathbf{X}(
bike\_safe\_distance \;\mathbf{U}\; in\_front\_bike
)
\]

\end{minipage}
}

  \caption{
  (a) Relational graph for the precondition of $\varphi$ shown in blue.
  (b) The ego (red) and NPC (black) paths.
  (c) Initial scene satisfies $behind\_bike$ and $bike\_safe\_distance$ ($in\_front\_bike$ pending).
  (d) During passing, $bike\_safe\_distance$ is violated.
  (e) After passing, $in\_front\_bike$ is satisfied.
  The NPC bike is marked with a red arrow and the ego is a blue car.
  }
  \label{fig:full_demo}
\end{figure*}

\section{Introduction}
\label{sec:intro}

The development of autonomous vehicles (AVs) is progressing rapidly with projections that tens of thousands of commercial AVs will be deployed in the coming years~\cite{goldman2025autonomous}. 
Like human drivers, these systems must adhere to rules
of safe driving, e.g., coming to a complete stop at a stop sign, maintaining a safe following distance, or maintaining a safe distance when overtaking another vehicle.
Researchers~\cite{woodlief2025sgsm,woodlief2025scene,toledo2025t4pc,sun2024redriver}
have developed promising methods for specifying such properties using pre and postconditions.
Here, the precondition defines the context for a driving rule, such as the spatial and temporal relationship of vehicles at an intersection, and the postcondition states allowable AV behavior in that context, like yielding the right-of-way.  

Validating an AV's adherence to such properties requires constructing a scenario that meets the precondition. For example, the scenario must show that as the \textit{ego} vehicle (the vehicle whose behavior is specified by the property) approaches the intersection other vehicles are already there, thereby establishing the context within which
the ego vehicle's behavior can be judged against the postcondition, i.e., whether it yields to the other vehicles or not.
Constructing such scenarios is the crux of testing AV agents, and it is typically performed using simulators like CARLA~\cite{Carla}.

Existing methods for performing such simulation-based property validation fall into two categories.
Techniques may develop a fixed set of scenarios~\cite{liu2025behavioralsafetyassessmentlargescale,10234383}
or leverage crash data to reconstruct test scenarios~\cite{bashetty2020deepcrashtestturningdashcamvideos,zhou2025efficientsafetytestingautonomous} to produce tests that meet the precondition.  However, there may be multiple ways to establish
the precondition, e.g., different timings and numbers of vehicles arriving at an intersection, and this prior work does not explore
this diversity.
In contrast, fuzzing using algorithmic~\cite{10.1145/3727875} 
or LLM-based approaches~\cite{10852505,  mei2025seekingcollideonlinesafetycritical} can produce large, and
potentially diverse sets of tests, but those tests may fail to
meet the particular precondition rendering them of little value for validating
a target property.

In this paper, we develop an automated approach to
\textit{Specification-based Testing for Autonomous Driving
Agents} (\stada).
\stada builds on \sceneflow~\cite{woodlief2025scene} specifications
that describe temporal relationships, using \ltlf~\cite{de2013linear}, between states of a system that
are described by spatial relationships among objects in a scene.
\Cref{fig:full_demo} shows a \sceneflow specification, $\varphi$,
of an overtaking maneuver of a \textit{non-player character} (NPC) -- a simulation entity not controlled by the agent under test. The precondition (shown in blue) 
requires that the ego vehicle is behind ($behind\_bike$) and
at a safe distance ($bike\_safe\_distance$) from a bike, and in the future
it is in front of the bike ($in\_front\_bike$).  These atomic propositions are defined using relational logic over 
a scene graph (SG) computed from sensor inputs, e.g., $behind\_bike \equiv ego.aheadOf \cap Bike \not= \emptyset$.
The test generation challenge is to configure simulation environments that bias the agent controlling the ego to meet this precondition, after which the post-condition can be checked, e.g., that a safe distance is maintained through the maneuver.

To achieve this, as depicted in \Cref{fig:full_demo}, \stada
analyzes the \ltlf structure of the specification and defines a set of
\textit{relational graphs} (RG) that contain distinct scenarios that meet the precondition (1.a).  These graphs encode requirements both on the initial state in the scenario (1.c) and on the trajectories through the environment established by that initial state.  \stada computes the space of trajectories consistent with the precondition (1.b) and selects a diverse subset of those for simulation.  The initial state and trajectories are used to generate simulation code that establishes the initial state (1.c) and then biases the ego's agent towards meeting intermediate (1.d) and final (1.e) states corresponding to the precondition.

We implemented \stada using the \scenic~\cite{Scenic} language to generate the initial scene and Python to generate CARLA~\cite{Carla} simulations. An evaluation using two different driving agents and comparing against three baseline techniques shows that \stada is cost-effective in automating the generation of diverse simulations that are consistent with temporal-logic driving properties.
It yields more than twice the coverage of the space of precondition-consistent behaviors than the best baseline technique and can match the coverage of those techniques in 6 times fewer simulations.
We describe the \stada approach in \Cref{sec:approach}, its
evaluation in \Cref{sec:evaluation}, and background and related work next.

\section{Related Work and Background}
\label{sec:related}
We briefly survey related work on coverage, test generation, and property specification and monitoring for AVs and provide relevant background on \scenic and \sceneflow. 
%\trey{re-org when sections are finalized}
% \trey{SGGs, specifying AV properties, SGSM/SceneFlow, Test Generation}

\subsection{Coverage and Test Generation for AVs}
The test adequacy problem aims to determine whether a test suite sufficiently exercises a given system under test (SUT), typically defined through a \textit{coverage criteria}.
Prior work in the AV space has measured coverage for system-level testing based on the number of \textit{scenarios of interest} that are covered~\cite{tang2023survey}, where the scenario universe is defined using features such as the road structure~\cite{tang2021collision} or vehicle interactions~\cite{tahir2021intersection}.
Specification-based testing evaluates an SUT with respect to its formal specifications to increase confidence in correctness~\cite{schlingloff2022specification,pecheur2009formal}.
Prior work on AV test coverage proposed Spatial Semantic Scene Coverage, which used specifications to define a coverage domain based on individual scenes~\cite{woodlief2024s3c}.
\stada extends this by supporting test generation and coverage of \textit{temporal} specifications.
% In this work, we extend this idea to include a rich notion of \textit{scenario coverage} based on \textit{temporal specifications}. \trey{probably need to distinguish between scene and scenario earlier in the paper}

% \se{check paragraph - revised/integrated two into 1}
Prior work has also proposed test generation techniques, including automated methods such as random~\cite{woodlief2024s3c} or coverage-based scenario generation~\cite{tang2021collision,tahir2021intersection}, and manual methods which encode specific scenarios of interest to ensure coverage of, e.g., prior crash scenarios~\cite{thorn2018framework}.
%\trey{I think it'd be more natural to introduce testing before coverage?}\se{I think is ok}
%Generating system-level AV test scenarios requires specifying not only the test environment, i.e., the physical environmental conditions of the test, but also the temporal elements, i.e., the configuration and actions of the other dynamic actors in the environment, e.g., the speed and driving behavior of the other cars.
%\scenic is a probabilistic scenario description language widely used to describe AV test cases~\cite{Scenic}. However, the expressive power of \scenic directly leads to complexity in leveraging this power to generate test cases relevant to the testing goals. To this end, ScenicNL~\cite{scenicNL} leverages large language models to translate crash reports into functional \scenic tests, automating prior manual work for creating tests based on crash reports.
%System-level AV tests require a description of the static environment---the stop sign and intersection---and of the dynamic elements---the behaviors of the other vehicles over time.
Tools such as \scenic, 
% which is 
a probabilistic scenario description language, are used to describe AV test cases capable of capturing rich scenes and dynamic behaviors~\cite{Scenic} and have been used to convert natural language crash scenarios into operable tests~\cite{scenicNL}. We built on this tool and use it as a baseline. 
%to \trey{need another example. verifai seems reasonable}~\cite{dreossi2019verifai}.

% Our technique is the first to combine a specification-based spatial-temporal notion of coverage with corresponding test generation, creating tests in \scenic that aim to directly exercise the AV's specifications.
While prior approaches measure coverage or generate scenarios, they do not leverage spatial-temporal specifications to systematically generate tests and measure coverage.

%\joy{removed SG from the approach, evaluation discusses that scene graph generator generates simulation trace without explicitly mentioning the scene graph. So commented out SG subsection from the background.}
% \subsection{Scene Graphs}

% \se{This remains disconnected -- do we need SGs here? How are we using them later?}
% \se{And note that after reading that section under approach, I am unsure where does SGs come from -- aren't SG really only used in evaluation?}
% A scene graph (SG) is a structured representation of the environment where the nodes in the graph correspond to the objects in the environment and the edges correspond to their spatial or semantic relationships~\cite{roadscene2vec,woodlief2024s3c}.
% A scene graph generator (SGG) builds a scene graph based on the current environment, either using sensor data~\cite{roadscene2vec,toledo2025monitoring} or, in simulation contexts, the simulator state~\cite{woodlief2025closing}.
% SGGs can be parameterized to capture the information needed for a given task~\cite{woodlief2024s3c}, and have been leveraged for collision prediction~\cite{roadscene2vec}, runtime monitoring~\cite{woodlief2025sgsm, woodlief2025scene}, and coverage analysis~\cite{woodlief2024s3c} in the AV context.

\subsection{Specifying AV Properties}
\label{sec:spec_av_props}
To systematically generate test inputs from specifications, AV safety requirements must be formalized in a logic that encodes environmental aspects using primitives that preserve their physical semantics.
% Recognizing this~\cite{fm4av_survey}, prior research has encoded portions of driving codes of several jurisdictions, including China~\cite{sun2022lawbreaker, sun2024redriver}, Europe~\cite{fm_german_dm}, and the United States~\cite{fm_usa_dm, woodlief2025sgsm}.
% %, and the Netherlands~\cite{fm_dutch_dm}.
% % in various logics.
% These specifications use temporal logics to express the behavior of the AV in response to the environment through time.

% \trey{this is currently structured as: prev subsection on SGs, independent par about related work on specs, par 2 on RFOL, par 3 on LTLf, par 4 on sceneflow. These need to be reworked to flow together better so the reader understands why we are introducing all of these---I don't think the takeaway is clear. I can take a stab at that later today, but if someone beats me to it, feel free.}

Prior work on runtime monitoring for AVs has used Relational First-Order Logic (RFOL)
% \trey{during my job talks/defense, I got some push back that this isn't actually \textit{first-order}, i.e., we don't use any part that would distinguish ``relational logic'' from RFOL}
 over scene graphs (SGs)~\cite{woodlief2025sgsm, woodlief2025scene} to express robust and diverse specifications.
The vertices of an SG represent entities in a scene, and the edges represent
spatial relationships between entities.
RFOL expressions describe sets of SG vertices, using operators $\{\cup,\cap,\setminus,\triangle\}$;
the relational image operator, $S.r$, which computes the set of 
vertices reachable using a designated relation, $r$, starting with
any vertex in a given set, $S$.
Resulting sets can be queried to
determine their size, $\lvert S \rvert$, and this can be used, for
example, to express non-emptiness, $\lvert S \rvert > 0 \equiv S \not= \emptyset$.

While RFOL captures spatial relationships at a single time instant, AV safety requirements also specify how these relationships evolve over time. To support this, Linear Temporal Logic over Finite Traces or \ltlf~\cite{de2013linear} is used
to express finite-horizon, discrete-time specifications that are well-suited for AV safety requirements.
\ltlf contains the temporal operators globally, $\mathbf{G}(\varphi)$, eventually, $\mathbf{F}(\varphi)$, next, $\mathbf{X}(\varphi)$, and until, $(\varphi_1)\mathbf{U}(\varphi_2)$. Prior work has demonstrated the utility of \ltlf to express safe driving requirements in runtime monitoring~\cite{woodlief2025sgsm, woodlief2025scene}.
This is enabled by the fact that an \ltlf specification translates to a deterministic finite automaton (DFA) which can be used to efficiently determine whether a trace meets the specification~\cite{fuggitti_ltlf2dfa}.

% In this work, we focus on the use of RFOL and \ltlf in the context of 
\sceneflow specifications~\cite{woodlief2025scene} combine atomic propositions (APs) built from RFOL over scene graphs, e.g., capturing that the ego's lane is controlled by a stop sign, and \ltlf, e.g., specifying the 
% that specifies the 
temporal sequence of stopping, waiting, and proceeding from the stop sign when clear.
\begin{figure*}[ht]
    \centering
    \includegraphics[width=0.95\textwidth]{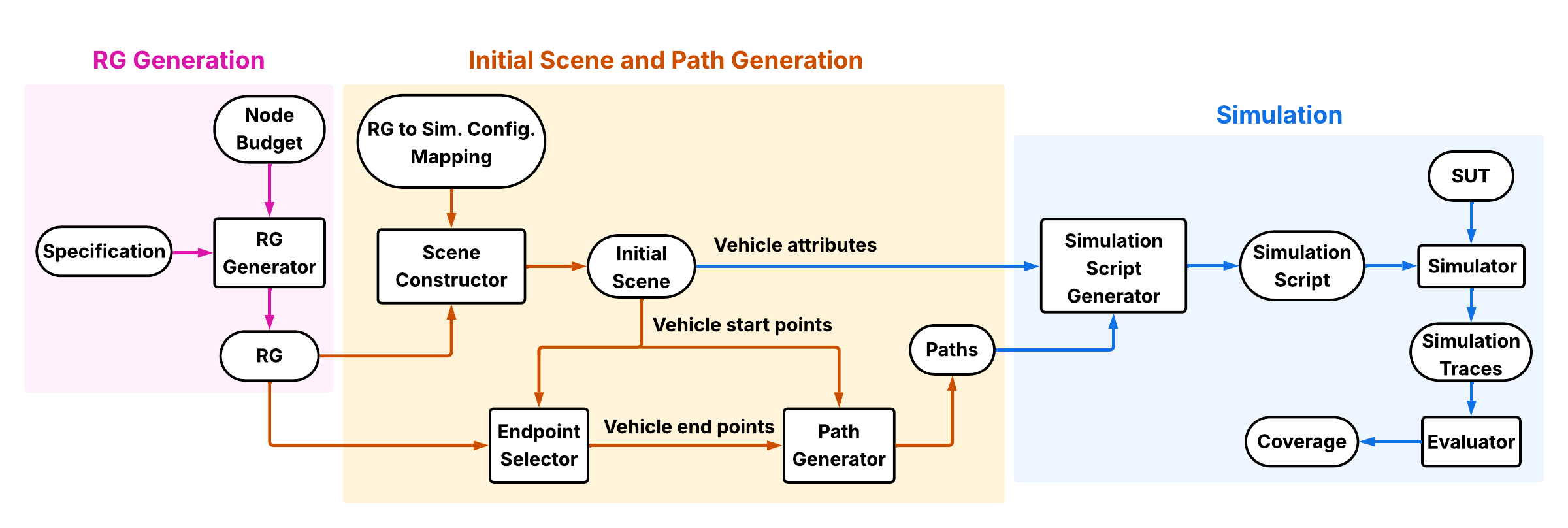}
    \vspace{-3mm}
    \caption{Overview of \stada.} 
    \label{fig:stada_overview}
\end{figure*}

In this work, in contrast to other test generation techniques, \stada leverages this structured specification framework for systematic test input generation. \stada uses the individual RFOL APs to constrain the space of \textit{scenes} to consider and the \ltlf to constrain the temporal evolution of the \textit{scenario} to guide test generation based on the specification.

\sceneflow specifications, $\varphi=\varphi_{pre}\rightarrow\varphi_{post}$, can be decomposed into precondition, $\varphi_{pre}$, and postcondition, $\varphi_{post}$, pairs.
% In this way, \sceneflow can evaluate whether the precondition was met using $\varphi_1$ and further evaluate the test itself using $\varphi$.
%Although no example is studied in this work, 
If a specification cannot be decomposed into this form, \stada uses the full specification $\varphi$ to bias test generation towards relevant scenarios.
Since $\varphi_{pre}$ describes a condition that is initially false and that becomes true at some point in a trace, it will typically not contain the $\mathcal{G}$ operator.
\stada assumes preconditions are $\mathcal{G}$-free in \Cref{sec:rg_generation}.

\section{Approach}
\label{sec:approach}

As a specification-based testing approach, \stada aims to \textit{cover} the different ways a system can satisfy the specification.
To achieve this for \ltlf preconditions, \stada must address the
fact that both the \ltlf formula and the RFOL expressions defining
atomic predicates (APs)
symbolically encode an enormous space of system behavior.

\subsection{Key Insight}

Consider the precondition expressing a vehicle to the left of or behind the $ego$, and in the future a car in front of it:
\[
\lvert \textit{ego.left} \cup \textit{ego.behind} \rvert > 0 \wedge F(\lvert \textit{ego.aheadOf} \cap Car \rvert > 0)
\]
\noindent where $ego$ is the singleton set defining the vehicle under
test, $\{ \textit{left}, \textit{behind}, \textit{aheadOf} \}$ are relations, and $Car$ is the set of cars. 
  This might be used to capture the notion of the $ego$ being passed
by a car, allowing for the position of the passing car to evolve from being behind,
then on the left, then in front, over time.
While capturing this intent, the precondition specification allows for an enormous
variety of traces to satisfy it.  For example, such sequences could forgo ever
being behind the $ego$ by just overtaking in the left lane, or being behind and passing
on the right, and there may be cars both behind and to the left.
The generality of RFOL and \ltlf in capturing behavior is its strength, but it
presents challenges in determining how many different \textit{configurations}
of the simulation are required for exploring the diversity present in the precondition.

To address this, \stada leverages the fact that the traces generated from a finite simulation are bounded.
Moreover, as described in~\Cref{sec:related},
 \ltlf behavior can be encoded as the finite traces accepted by 
a deterministic finite state machine.  This scoping of behavior allows \stada to
define the structure of disjoint configurations that collectively cover the behavior
of an \ltlf precondition.
These configurations can be used to generate initial scenes of simulations and  paths for establishing 
conditions that aim to drive those simulations towards the goal of meeting
the preconditions.

\subsection{Approach Overview}

\stada consists of three modules as depicted in Figure \ref{fig:stada_overview}. The inputs to the framework consist of (i) the precondition of a safe-driving specification expressed in \ltlf, and (ii) a \textit{node budget} defining the number of entities for each node type referenced in the specification. 
% \se{not sure about this last piece, how does one determine what is the minimum number?}\joy{the minimum number is a complementary input along with the LTL, different node budget with the same LTL shows different intended scenario}. 
% \matt{Is it possible to compute the minimum budget given the structure of the LTL?  If so then the budget can be an option that allows for finer degrees of coverage.  The current explanation does not motivate the need for budgets nor does it talk about the consequences of making a bad choice.}\joy{current approach does not have such algorithm, we can mention it as a future direction.}
A minimum budget ensures coverage of precondition scenarios, while increasing the budget allows for more and more diverse tests to be generated.
The process begins with the generation of a set of relational graphs (RGs) by the \textcolor{magenta}{RG Generation} module, each of which represents a distinct system behavior of satisfying the precondition. For each such RG, the \textcolor{Bittersweet}{Initial Scene and Path Generation} module generates a static initial scene that satisfies the initial conditions, including spatial, topological, and placement constraints, and then generates feasible paths for the ego and all other relevant entities in the scene, i.e., the NPCs.
The \textcolor{blue}{Simulation} module integrates the initial scene and paths to run simulations using the SUT controlling the ego and the Evaluator
checks the resulting simulation traces against the \ltlf, which enables postcondition evaluation upon precondition satisfaction.  

\subsection{RG Generation}
\label{sec:rg_generation}

Figure \ref{fig:stada_overview} first shows 
the \textcolor{magenta}{RG Generation} module, which analyzes an $LTL_f$ specification to identify all of the ways its precondition can be satisfied. It decomposes the specification into configurations, which are distinct sets of choices that together cover all scenarios consistent with the precondition.

The generator begins by preprocessing the $LTL_f$ formulas. First, operators like implication are converted to negation, conjunction, and disjunction. Second, disjunctions $\varphi_1 \lor \varphi_2$ are split into three mutually exclusive cases: (i) $\varphi_1 \land \neg \varphi_2$, (ii) $\neg \varphi_1 \land \varphi_2$, and (iii) $\varphi_1 \land \varphi_2$. This results in a set of conjunction-only formulas, which the RG generator processes independently. 

The RG generation aims to produce a set of configurations, each of which is defined as a relational graph: $RG = (V, E, L_E)$,
where $V$ contains nodes corresponding to the entity types referenced in the specification, 
%\se{what SG? where does the SG comes from? is it an input to the approach?}\joy{removed SG} 
and $E \subseteq V \times V$ are edges. The labeling function $L_E : E \mapsto R@T$ maps edges to a relation $R$ and a temporal operator $T \in \{ \mathbf{I, X, F, U_l, U_r} \}$. 
Here, we introduce $\mathbf{I}$ to define constraints for initial states, the unary temporal operators $\mathbf{X, F}$ annotate edges with constraints, and the binary operator $\mathbf{U}$ is modeled by pairs of edges with constraints for the left and right operands.
For example, an edge labeled $R@\mathbf{F}$ from $s$ to $t$ implies that eventually $t \in s.R$. Binary operators like ``Until'' ($\mathbf{U}$) are modeled via pairs: $R_1@\mathbf{U_l}$ and $R_2@\mathbf{U_r}$ represent $(t_1 \in s_1.R_1) \mathbf{U} (t_2 \in s_2.R_2)$. If a temporal operator is missing from an AP, it defaults to $\mathbf{I}$ (Initially).

% \matt{I think the definition of LTL from above should be in background.  Then we should say here that we introduce I to define constraints for initial states, the unary temporal operators annotate edges with constraints, and the binary operator is modeled by pairs of edges with constraints for the left and right operands.}

% Figure \ref{fig:rg} illustrates an RG for the \ltlf specification presented in its caption. \se{are we using the specific temporal operators introduced here somewhere else? where? I am trying to decide whether this is needed for the following explanations or not} \trey{I find this a bit hard to follow---between $U$ requiring special handling and $I$ being introduced here, I don't think this is obvious unless you've seen the implementation already. I don't have an immediate fix, but something to think on.}

The process starts with a relational graph G containing only the ego node and iteratively refines it to satisfy each $AP_i$ in the specification. For each $AP_i$, as depicted in Algorithm~\ref{alg:generate-cand}, the algorithm first decomposes $AP_i$ into a set of $(src_j, rel_j, \tau)$ tuples using all the relational image operators referenced in the RFOL expression and extends G with additional nodes
of type $\tau$ up to the predefined node budget, where $\tau$ denotes the vertex type of elements in the set computed by the RFOL expression in $AP_i$.
% \trey{I don't think this is correct, and is making it a bit difficult for me to follow. the algorithm as written takes in a single AP as an argument, so it does not ``iteratively ... each AP'' --- there has to be a loop above this that goes through the APs itself? I see the ``where" clause, but that doesn't to me jive with the output here} \joy{Yes, there is an outer loop that does the preprocessing, splits the LTL, run \Cref{alg:generate-cand}, checks if the RG set satisfies the APs and consistency and then iterates for the next AP. For space constraints, I did not show the outer loop and named the algorithm \textbf{Candidate RG Generation for an AP}. I have reorganized the two lines here and mentioned \Cref{alg:generate-cand} after mentioning that we iterate for each AP.} 
It then enumerates all ordered tuples $(S_1, \dots, S_m)$ where each $S_j$ is a subset of the candidate node set $\mathcal{D}$, i.e., it explores the Cartesian product $\mathcal{P}(\mathcal{D})^m$ and adds edges $(src_j, rel_j, v)$ for all $v \in S_j$, thereby generating candidate RG set for $AP_i$. Only graphs satisfying consistency checks (e.g., a vehicle cannot be simultaneously ahead of and behind another) are passed to the next iteration. 
%\se{Are we specifying the consistency check somewhere? Are they simple newtonian-like constraints? Or are they an input to the RG generation as well?}\joy{no we are not specifying because they are simple newtonian-like spatial constraints.}
Failing to find at least one satisfiable graph at any iteration, the algorithm returns no solution, indicating that the \ltlf formula is not satisfiable within the given node budget. The final output of this module is a set of structurally unique RGs, obtained after removing isomorphic duplicates, each of which satisfies the precondition in a distinct way.

\begin{algorithm}[t]
\caption{Candidate RG Generation for an AP}
\label{alg:generate-cand}
\small
\begin{algorithmic}[1]
\STATE \textbf{Input:} Initial graph $G$, Atomic Proposition $AP_i$
\STATE \textbf{where} $AP_i = \{(src_j, rel_j, \tau) \mid 1 \le j \le m\}$
\STATE \textbf{Output:} Set of candidate RGs $\mathcal{C}$

\STATE $n \leftarrow \textit{NodeBudget}(\tau) - |\{v \in G \mid \text{type}(v) = \tau\}|$
\STATE Add $n$ new nodes of type $\tau$ to $G$
\STATE $\mathcal{D} \leftarrow \{v \in G \mid \text{type}(v) = \tau\}$  
\STATE $\mathcal{C} \leftarrow \emptyset$

% \FORALL{combinations $\langle S_1, \dots, S_m \rangle$ where $S_j \subseteq \mathcal{D}$}
\FORALL{$(S_1, \dots, S_m) \in \mathcal{P}(\mathcal{D})^m$} %\Comment{Powerset bc?}
    \STATE $G' \leftarrow \text{copy of } G$
    \FOR{$j = 1$ \TO $m$} %\Comment{Iterate X}
        \FORALL{$v \in S_j$} %\Comment{Iterate Y}
            \STATE Add edge $(src_j, rel_j, v)$ to $G'$
        \ENDFOR
    \ENDFOR
    \STATE Remove all $v \in \mathcal{D}$ where $\text{deg}(v) = 0$ in $G'$  
    \STATE $\mathcal{C} \leftarrow \mathcal{C} \cup \{G'\}$
\ENDFOR

\RETURN $\mathcal{C}$
\end{algorithmic}
\end{algorithm}

\subsection{Initial Scene and Path Generation}
\label{sec:path_gen}
As depicted in Figure \ref{fig:stada_overview}, 
the Scene Constructor takes a relational graph (RG) and its corresponding simulation mappings to generate a simulation scene.  By leveraging a scene construction tool (e.g., \scenic, OpenSCENARIO~\cite{OpenSCENARIO}), it then instantiates an initial static scene in a simulator's native language. 
%such that it satisfies all initial constraints of the RG, i.e., it transforms high-level constraints into concrete spatial coordinates. 
Formally, using the given mapping $\mathcal{M}$ and the $RG$, the Scene Constructor produces a simulator scene $\Sigma$ and instantiates it using a scene construction tool $\mathcal{E}$, getting an initial scene $\mathcal{X}$: 
%\begin{equation}$
$\quad \mathcal{M} : RG \rightarrow \Sigma
\quad \text{and} \quad
\mathcal{E} : \Sigma \rightarrow \mathcal{X}$
%\label{eq:rg_mapping}
%\end{equation}
This initial scene captures the poses of the ego as well as the NPCs, and the generated coordinates serve as markers for the subsequent path generation. 

Once the initial poses are identified, the module generates routes that are consistent with the scene. This is achieved through a two-stage process designed to bias trajectories toward satisfying a target RG. 
First, the module identifies the trajectories' endpoints for the vehicles. It binds RG concepts like lane and intersection nodes to concrete road entities in the simulator’s road network. In cases where the binding cannot be unequivocally established, the module defaults to a valid entity. Then, to determine the final destination for each vehicle, it samples a set of simulator waypoints within an initial radius $r$ of the starting point, it filters that set based on the RG edges (e.g., to the side, behind). If the filter returns an empty set, the radius $r$ is increased, and the space is resampled until a feasible set of endpoints is obtained from which one is randomly selected as the final endpoint.

Second, the module constructs a dense waypoint graph onto the scene topology where waypoints are nodes and legal transitions are edges \cite{woodlief2025closing}. This higher resolution representation enables the expression of fine-grained behaviors, such as mid-lane changes or overtaking. Then, the module maps the starting and ending positions for the ego and NPCs on this graph to begin trajectory planning.
It uses the K-shortest simple path algorithm~\cite{10.1287/mnsc.17.11.712} to find $k$ feasible paths. To increase structural diversity and coverage of maneuvers such as lane changes and positional reordering, the module employs  greedy selection, iteratively choosing  $p^\star$ from the set of candidates $\mathcal{O}$ that maximizes its average distance from the set of already-selected paths $\mathcal{L}$: $
p^\star = \arg\max_{p \in \mathcal{O} \setminus \mathcal{L}}
\frac{1}{|\mathcal{L}|}
\sum_{q \in \mathcal{L}} D(p,q)$
%\end{equation}
where $D(p,q)$ is the average Euclidean distance between the points of  $p$ and $q$.
 % $D(p,q)$ is the average Euclidean distance between the points of the paths.
%\begin{equation}
%D(p,q) =
%\frac{1}{|P(p)|}
%\sum_{x \in P(p)}
%\min_{y \in P(q)} \|x - y\|_2,
%\end{equation}
% where $P(p)$ and $P(q)$ denote the sets of points along paths $p$ and $q$, respectively. 
% \trey{Since the use of P was commented out, I removed the prose too.}

\subsection{Simulation}
\label{sec:simulation}
This module receives the vehicle attributes (e.g., vehicle models) and the paths for all vehicles from the \textcolor{Bittersweet}{Initial Scene and Path Generation} module, and, using the Simulation Script generator, produces an executable simulation script that spawns the vehicles at the starting positions of their paths and assigns driving controllers to them. The ego is controlled by the system under test (SUT), while all NPCs are driven using driving agents on autopilot.
% \trey{From an approach level, the NPCs could be driven by approach-specified controllers that are much more robust than what we explore here. In the next paragraph there are statements like ``some autopilots''---it is unclear what the set of autopilots we are discussing over is.}

We assume that the only externally controllable parameter for the SUT is the route.
%, whereas for autopilot-controlled NPCs, both the speed and the route are tunable. 
For the NPC, the module can control steering angle, throttle, speed, and the route. Since the NPC speed is a key variable that can directly influence the precondition satisfaction, and keeping it constant for all specifications would be insufficient to generalize effectively, we leverage the fact that most of the traffic laws tend to trigger if the ego and the NPCs are within close proximity.
% So,
As such, we dynamically update the speed of the NPCs at each simulation step by decreasing it if the NPC is ahead of the ego and increasing it if it is behind the ego, so that they tend to stay closer to each other. 
% The relative position of the NPC with respect to the ego can be determined using Equation \ref{eq:signed_separation}]
% \begin{equation}
% s \;=\; (\mathbf{p}_{\text{ego}} - \mathbf{p}_{\text{npc}})^\top \mathbf{f}_{\text{npc}},
% \label{eq:signed_separation}
% \end{equation}
% where $\mathbf{p}_{\text{ego}}$ and $\mathbf{p}_{\text{npc}}$ denote the positions of the ego and NPC vehicles, respectively, and $\mathbf{f}_{\text{npc}}$ is the unit forward direction vector of the NPC.
% Concretely, $s < 0$ indicates that the NPC is placed ahead of the ego along its forward direction, while $s \ge 0$ indicates that it is behind the ego. 
Additionally, the amount of speed change is proportional to the distance between the vehicles. We use the longitudinal separation between the vehicles instead of the Euclidean distance because lateral distance should not influence speed change. 
\begin{equation}
v_{\text{npc}}(d)
=
\operatorname{clip}\!\left(
v_0 + (\alpha \mathbf{1}_{d\ge0} - \beta \mathbf{1}_{d<0})\, d,
\, v_{\min},\, v_{\max}
\right)
\label{eq:npc_speed}
\end{equation}
where $d$ represents the longitudinal deviation, $v_0$ is the nominal NPC speed, $\alpha$ and $\beta$ are proportional gain parameters, and $v_{\min}$ and $v_{\max}$ bound the speed within safe limits. This formulation ensures a smooth, bounded as well as a distance-proportional speed adaptation.

Using the selected agents, paths, and dynamically adjusted speeds, the \textcolor{blue}{Simulation} module executes the simulation script and records the trace using a scene graph generator.

Finally, the Evaluator checks the traces against the \ltlf specifications to produce coverage metrics. % formulated below.
%\se{maybe one more sentence about how it does it? } \joy{We can add something like by converting it into dfas and then using the simulation trace to evaluate it. But I was thinking that as this is an abstract evaluator, it may have a completely different mechanism than SceneFlow. Should we add more? or keep the \textit{how} abstract that the evaluator could have a completely different approach to evaluate}
%\textbf{Coverage Metrics:}
%\noindent
% \se{Just in case it was lost in my previous comments I will repeat it here: I think this section is out of place, is not just about this study, the definition of the coverage metrics could be part of the approach under Simulation}\joy{I will move this to approach, but do you think it will be ok if I refer $\varphi_3$ and $\varphi_4$ examples in it?}
Disjunctions in an \ltlf define different ways that a precondition may be satisfied, and the presence of multiple forms of disjunction in a precondition induces a combinatorial space of possible configurations to be explored during testing. Conceptually, a configuration defines a combination of choices for each disjunctive fragment of a precondition, e.g., the left or right operands of a $\vee$ in \ltlf or a $\cup$ in RFOL.

The coverage metrics are defined in increasing order of coarseness. The finest-grained metric, $cov_1$, measures how many distinct configurations are covered across all simulation traces of a particular treatment.
Another metric $cov_2$ is applicable to a specific class of \ltlf specifications in which multiple APs present in an earlier timestamp are flipped in a later timestamp.
We define a \textit{oneflip} configuration as one in which exactly one AP is flipped, and $cov_2$ specifies how many of these \textit{oneflips} are covered out of all possible \textit{oneflip} cases.
This approach is analogous to Modified Condition/Decision Coverage (MC/DC)~\cite{Chilenski1994ApplicabilityOM} in structural testing, as it ensures each AP independently affects the outcome. This isolation is critical for detecting edge-case bugs that might be masked if multiple APs always change simultaneously.
Finally, $cov_3$ is a binary coverage metric that checks whether at least one of the distinct configurations mentioned in $cov_1$ is covered.
\begin{table*}[!t]
\centering
\setlength{\tabcolsep}{4pt}
\renewcommand{\arraystretch}{1.25}
\newcolumntype{Y}{>{\raggedright\arraybackslash}X}

\captionof{table}{Preconditions of the \sceneflow specifications~\cite{woodlief2025scene} in natural language and the corresponding \ltlf{}s.}
\label{tab:rules}

\begin{tabularx}{\textwidth}{@{}r Y@{}}
\toprule

% φ0
$\varphi_0$ & \textbf{If another vehicle arrives at an uncontrolled intersection before the ego and both are later present at the intersection simultaneously,} then ... \\
& \footnotesize
\begin{tabular}[t]{@{}l@{}}
$(atInter(veh, \jay) \wedge \neg atInter(ego, \jay) \wedge hasStop(ego)) \wedge
\mathbf{X}(atInter(veh, \jay) \wedge atInter(ego, \jay))$
\end{tabular} \\
\midrule

% φ1
$\varphi_1$ & \textbf{If the ego and another vehicle approach an intersection and, in the next step, are at the intersection with the other to the right,} then ... \\
& \footnotesize
$(\neg atInter(veh,\jay) \wedge \neg atInter(ego,\jay) \wedge hasStop(ego) \wedge hasStop(veh)) \wedge \mathbf{X}(atInter(veh,\jay) \wedge atInter(ego,\jay) \wedge toRightOf(ego,veh))$ \\
\midrule

% φ1'
$\varphi_1'$ & \textbf{If the ego and another vehicle approach an intersection and are at the intersection at some future time with the other to the right,} then ... \\
& \footnotesize
$(\neg atInter(veh,\jay) \wedge \neg atInter(ego,\jay) \wedge hasStop(ego) \wedge hasStop(veh)) \wedge \mathbf{F}(atInter(veh,\jay) \wedge atInter(ego,\jay) \wedge toRightOf(ego,veh))$ \\
\midrule

% φ2
$\varphi_2$ & \textbf{If an emergency vehicle with lights on is present at a signaled intersection while the ego is entering there,} then ... \\
& \footnotesize
\begin{tabular}[t]{@{}l@{}}
$\neg atInter(ego) \wedge \mathbf{X}(atInter(veh, \jay) \wedge hasEmergencyLights(veh) \wedge atInter(ego, \jay) \wedge notEqual(veh, ego))$
\end{tabular} \\
\midrule

$\varphi_3$ & \textbf{If the ego is following another vehicle while the lead vehicle is moving,} then ... \\
% ego should not continue following too closely. \\
& \footnotesize
\begin{tabular}[t]{@{}l@{}}
$\neg (tooClose(ego, veh) \wedge sameLane(veh, ego) \wedge behind(ego, veh)) \wedge \neg stopped(veh) \wedge$\\
$\qquad\qquad\mathbf{X} (tooClose(ego, veh) \wedge sameLane(veh, ego) \wedge behind(ego, veh) \wedge \neg stopped(veh))$ 
\end{tabular} \\
\midrule

% φ4 <- φ8
$\varphi_4$ & \textbf{If the ego is following an emergency vehicle with sirens on,} then ... \\
% ego should not continue following too closely. \\
& \footnotesize
\begin{tabular}[t]{@{}l@{}}
$\neg (tooCloseToEmergency(ego, veh) \wedge sameLane(veh, ego) \wedge behind(ego, veh)) \wedge isEmergencyVehicle(veh) \wedge$\\
$\qquad\qquad\mathbf{X} (tooCloseToEmergency(ego, veh) \wedge sameLane(veh, ego) \wedge behind(ego, veh) \wedge isEmergencyVehicle(veh))$
\end{tabular} \\
\midrule

% φ5
$\varphi_5$ & \textbf{If the ego overtakes a bicycle,} then ... \\
& \footnotesize
\begin{tabular}[t]{@{}l@{}}
$(behind(ego, bike) \wedge safeDistance_D(ego, bike)) \wedge
\mathbf{F}(behind(bike, ego))$
\end{tabular} \\
\midrule

% φ6
$\varphi_6$ & \textbf{If the ego intends to move into the opposing lane to overtake another vehicle,} then ... \\
& \footnotesize
\begin{tabular}[t]{@{}l@{}}
$(behind(ego, veh) \wedge opposingClear(ego, \ell) \wedge onlyIn(ego, \ell)) \wedge
\mathbf{F}((\mathit{front}(ego, veh) \wedge onlyIn(ego, \ell)))$
\end{tabular} \\
\midrule

% φ7
$\varphi_7$ & \textbf{If the ego enters an intersection from a given lane,} then ... \\
& \footnotesize
\begin{tabular}[t]{@{}l@{}}
$onlyIn(ego, \ell_1) \wedge \mathbf{X}(\mathit{fullyInInter}(ego)) \wedge
\mathbf{X}(\mathbf{X}((\mathit{fullyInInter}(ego) \wedge \neg onlyIn(ego, \ell_2)) ~\mathbf{U}~ onlyIn(ego, \ell_2)))$
\end{tabular} \\

\bottomrule
\end{tabularx}
\end{table*}

% \input{scene_flow_prop_table}

% ---------------- Table B: Results by method ----------------
%\begin{table}[!t]
%\centering
%\small
%\setlength{\tabcolsep}{2pt}     % tighter columns
%\renewcommand{\arraystretch}{1.15}
%\newcolumntype{C}{>{\centering\arraybackslash}X}
%\caption{Rule-wise hit rate across methods}
%\label{tab:rq1}
%\begin{tabularx}{\columnwidth}{@{}r C C C C@{}}
%\toprule
%\textbf{\#} &
%\makecell{\textbf{Our}\\\textbf{Method}} &
%\makecell{\textbf{ScenicNL}\\\textbf{(zero-shot)}} &
%\makecell{\textbf{ScenicNL}\\\textbf{(few-shot)}} &
%\makecell{\carlabase} \\
%\midrule
%$\varphi_0$ & 100\% & — & — & 0\% \\
%$\varphi_1$ & 13.33\% & — & — & 0\% \\
%$\varphi_2$ & 20.00\% & — & — & 0\% \\
%$\varphi_3$ & 71.76\% & — & — & 2.5\% \\
%$\varphi_4$ & 56.47\% & — & — & 0\% \\
%$\varphi_5$ & 95.00\% & — & — & 0\% \\
%$\varphi_6$ & 5.00\% & — & — & 0\% \\
%$\varphi_7$ & 40.00\% & — & — & 5\% \\
%\bottomrule
%\end{tabularx}
%\end{table}

\section{Evaluation}
\label{sec:evaluation}
To evaluate the utility of \stada in test input generation for autonomous driving agents, we structure our experiments around the following two research questions:
\begin{enumerate}[label=\textbf{RQ\arabic*:}, leftmargin=*, itemsep=0.5em]
  \item \label{rq:hit_rate}
  How effective is \stada at generating simulations that cover the space of scenarios in the preconditions of the \ltlf specifications? 
  \item \label{rq:coverage}
  How efficient is \stada in covering such scenarios? 
\end{enumerate}

\subsection{Design}
\label{sec:design}
To answer these questions, we require: a set of driving properties in \ltlf, a scene construction tool, a driving simulator, autonomous driving agents as the system under test (SUT), and an evaluator to measure coverage.

For properties, we use the \ltlf specifications defined in \sceneflow \cite{woodlief2025scene} derived from Virginia Driving Laws. 
These specifications, shown in \Cref{tab:rules}, express complex traffic situations, and the language has been shown capable of expressing 110 of 114 (96\%) properties in the Regulation of Traffic of the Virginia driving code (Chapter 8) that are applicable to the autonomous driving agent~\cite{woodlief2025scene}.

We use \scenic~\cite{Scenic} as the scene construction tool in \stada. We build an RG-to-\scenic mapping for each potential relation of the RG. In some cases, multiple edges together instantiate a single code block, such as \texttt{isInIntersection} combined with edges specifying the type of intersection control (e.g., traffic lights, stop signs).
The \texttt{require} statement in \scenic enforces a hard constraint on the generated scene. 
For example, an edge in the RG ($ego$, $near$, $car_1$) with $time =I$ can be expressed in \scenic by the statement \texttt{require (10 < distance from $ego$ to $car_1$ <= 16)} to enforce that in the initial scene the $ego$ vehicle must be within 10 to 16 meters of $car_1$, matching the precise semantics of the $near$ relation. 
Distance and angular relationships in this mapping follow the definitions and physical semantics established in prior work~\cite{woodlief2025closing}.
As shown in \Cref{sec:path_gen}, given a list of parameters such as town, weather, etc., and this mapping $\mathcal{M}$, the \textit{Initial Scene and Path Generation} module iterates through each edge of the RG and incrementally builds a complete \scenic program $\Sigma$.

We use the popular CARLA 0.9.15~\cite{Carla} as the simulator, which has diverse road networks, realistic physics simulations, and compatibility with \scenic. 
For our experiments,
we use the optimized version of \textit{Town10HD} as the map since it has all the necessary road topologies to meet the preconditions in \Cref{tab:rules}, and \textit{ClearNoon} as the weather running at \textit{20Hz}. 
%We select \textit{Town10HD} because the targeted specifications are readily applicable to an urban environment, and this town has all necessary road topologies to meet the preconditions in \Cref{tab:rules}.
%However, our framework is applicable to any CARLA town.

%We study two SUTs.
Following previous studies~\cite{woodlief2025sgsm,woodlief2025scene}, we choose Interfuser and Transfuser++ \cite{10.1109/TPAMI.2022.3200245} as our two SUTs since they were the best performing agents in CARLA Leaderboard 1 and 2  ~\cite{carla_leaderboard}, with available code and model weights.  The NPCs are controlled by the CARLA Basic Agent, which allows them to follow a predefined path independent of traffic law.

% Prior researchers introduced 
\sceneflow~\cite{woodlief2025scene} can determine the satisfaction of an \ltlf specification over a set of finite traces generated by the CARLA Scene Graph Generator~\cite{woodlief2025closing},
% from CARLA simulations,
which we use as the Evaluator
% of our framework mentioned in
from \Cref{sec:simulation}.
Using the decomposition of $\varphi=\varphi_{pre}\rightarrow\varphi_{post}$ discussed in \Cref{sec:spec_av_props}, \sceneflow can evaluate whether the precondition was met using $\varphi_{pre}$ and further evaluate the test itself using $\varphi$.

\subsection{Treatments}
\label{sec:treatments}
To the best of our knowledge, no existing approach can generate test inputs for autonomous driving agents directly from \ltlf. Only a few tools are capable of generating test inputs or scenarios from a natural language description, but most lack publicly available or fully functional implementations \cite{10852505,mei2025seekingcollideonlinesafetycritical}.
% \se{make sure to discuss those in related work and point back to them in here}
Nonetheless, we present 4 treatments for our experiments: two stochastic-placement baselines, a state-of-the-art \scenic scenario generation technique, and \stada.

To target each RG, we run CARLA by spawning the ego and the same number of NPCs as specified in the \stada RG at random locations. 
Each such configuration is executed the same number of times as \stada, with the ego controlled by the SUT and NPCs by the CARLA Basic Agent.
We denote this baseline as \carlabase.
We create another similar baseline with 10 times more vehicles than \carlabase.
%in the RGs. 
%For example, if an RG has one ego and a generic car, this baseline spawns one ego and 10 generic cars.
%Similarly, for an RG with one ego and one emergency vehicle, this baseline is executed with 10 emergency vehicles at random locations using the same agent selection policy as \carlabase.
We include this \carlaTenX baseline to explore how well-resourced
% any
brute force exploration performs.

We use ScenicNL~\cite{scenicNL}, which generates \scenic code from natural language (NL) scenario descriptions as a third baseline.
For each \ltlf a manually written scenario description is provided to ScenicNL as a prompt along with 3 example \scenic programs and their NL descriptions.
These few-shot examples collectively contain at least one instance of all boolean and temporal operators and specifiers required for expressing any of our target specifications. We use \textit{GPT-4-0613} as the LLM backbone and prompt ScenicNL once for each \ltlf to generate the \scenic code.
% ScenicNL creates a set of behaviors and assigns one behavior to each NPC  which is controlled by the \scenic autopilot. 

For \stada, we generate two initial scenes for each RG and execute each initial scene with two different paths, generating 4 simulations for each RG. 
Across the 20 RGs (excluding $\varphi_1'$) for the specifications in \Cref{tab:rules}, this gives a total of 80 simulations.
We choose to generate two initial scenes and two paths per scene to minimally validate our framework's ability to produce multiple valid and diverse instantiations at the initial scene and path generation stages.

We execute each baseline 80 times using randomly generated paths. However, ScenicNL produces only 3 compilable codes out of the 8 \ltlf specifications. Here, if \stada generates $n$ simulations for $\varphi_i$, we run the corresponding ScenicNL program $n$ times as well. In total, this results in 36 simulations, and the remaining 44 are considered empty.

\subsection{\textit{RQ1: Coverage Results}}

\begin{table*}[t]
\centering
\small
\setlength{\tabcolsep}{2.0pt}
\renewcommand{\arraystretch}{1.02}

\begin{adjustbox}{max width=\textwidth}

\begin{tabular}{c||ccc|ccc|ccc|ccc||ccc|ccc|ccc|ccc}
\toprule
\multirow{3}{*}{\thead{Spec.\\(Feasible \\ RG/ \\ Total RG)}}
& \multicolumn{12}{c||}{\textbf{TransFuser}}
& \multicolumn{12}{c}{\textbf{InterFuser}} \\
\cmidrule(lr){2-13} \cmidrule(lr){14-25}

\multicolumn{1}{c||}{}
& \multicolumn{3}{c|}{\thead{\carlabase}}
& \multicolumn{3}{c|}{\thead{\carlaTenX}}
& \multicolumn{3}{c|}{\thead{ScenicNL\\(few-shot)}}
& \multicolumn{3}{c||}{\thead{\stada}}
& \multicolumn{3}{c|}{\thead{\carlabase}}
& \multicolumn{3}{c|}{\thead{\carlaTenX}}
& \multicolumn{3}{c|}{\thead{ScenicNL\\(few-shot)}}
& \multicolumn{3}{c}{\thead{\stada}} \\

\cmidrule(lr){2-4} \cmidrule(lr){5-7} \cmidrule(lr){8-10} \cmidrule(lr){11-13}
\cmidrule(lr){14-16} \cmidrule(lr){17-19} \cmidrule(lr){20-22} \cmidrule(lr){23-25}

\multicolumn{1}{c||}{}
& cov$_1$ & cov$_2$ & cov$_3$
& cov$_1$ & cov$_2$ & cov$_3$
& cov$_1$ & cov$_2$ & cov$_3$
& cov$_1$ & cov$_2$ & cov$_3$
& cov$_1$ & cov$_2$ & cov$_3$
& cov$_1$ & cov$_2$ & cov$_3$
& cov$_1$ & cov$_2$ & cov$_3$
& cov$_1$ & cov$_2$ & cov$_3$ \\
\midrule

$\varphi_0$ (1/1)
& 0/1 & 0/1 & 0/1
& \textbf{1/1} & \textbf{1/1} & \textbf{1/1}
& 0/1 & 0/1 & 0/1
& \textbf{1/1} & \textbf{1/1} & \textbf{1/1}
& 0/1 & 0/1 & 0/1
& \textbf{1/1} & \textbf{1/1} & \textbf{1/1}
& 0/1 & 0/1 & 0/1
& \textbf{1/1} & \textbf{1/1} & \textbf{1/1} \\

\rowcolor{lightgray}
$\varphi_1$ (1/1)
& 0/1 & 0/1 & 0/1
& 0/1 & 0/1 & 0/1
& 0/1 & 0/1 & 0/1
& 0/1 & 0/1 & 0/1
& 0/1 & 0/1 & 0/1
& 0/1 & 0/1 & 0/1
& 0/1 & 0/1 & 0/1
& 0/1 & 0/1 & 0/1 \\

$\varphi_1'$ (1/1)
& 0/1 & 0/1 & 0/1
& 0/1 & 0/1 & 0/1
& 0/1 & 0/1 & 0/1
& \textbf{1/1} & \textbf{1/1} & \textbf{1/1}
& 0/1 & 0/1 & 0/1
& 0/1 & 0/1 & 0/1
& 0/1 & 0/1 & 0/1
& \textbf{1/1} & \textbf{1/1} & \textbf{1/1} \\

\rowcolor{lightgray}
$\varphi_2$ (1/1)
& 0/1 & 0/1 & 0/1
& 0/1 & 0/1 & 0/1
& 0/1 & 0/1 & 0/1
& \textbf{1/1} & \textbf{1/1} & \textbf{1/1}
& 0/1 & 0/1 & 0/1
& 0/1 & 0/1 & 0/1
& 0/1 & 0/1 & 0/1
& \textbf{1/1} & \textbf{1/1} & \textbf{1/1} \\

$\varphi_3$ (4/7)
& 1/4 & 1/3 & \textbf{1/1}
& 2/4 & 2/3 & \textbf{1/1}
& 1/4 & 0/3 & \textbf{1/1}
& \textbf{4/4} & \textbf{3/3} & \textbf{1/1}
& 2/4 & 2/3 & \textbf{1/1}
& 1/4 & 1/3 & \textbf{1/1}
& 0/4 & 0/3 & 0/1
& \textbf{4/4} & \textbf{3/3} & \textbf{1/1} \\

\rowcolor{lightgray}
$\varphi_4$ (4/7)
& 0/4 & 0/3 & 0/1
& 1/4 & 1/3 & \textbf{1/1}
& 0/4 & 0/3 & 0/1
& \textbf{3/4} & \textbf{3/3} & \textbf{1/1}
& 0/4 & 0/3 & 0/1
& 2/4 & 2/3 & \textbf{1/1}
& 0/4 & 0/3 & 0/1
& \textbf{3/4} & \textbf{3/3} & \textbf{1/1} \\

$\varphi_5$ (1/1)
& 0/1 & 0/1 & 0/1
& 0/1 & 0/1 & 0/1
& 0/1 & 0/1 & 0/1
& \textbf{1/1} & \textbf{1/1} & \textbf{1/1}
& 0/1 & 0/1 & 0/1
& 0/1 & 0/1 & 0/1
& 0/1 & 0/1 & 0/1
& \textbf{1/1} & \textbf{1/1} & \textbf{1/1} \\

\rowcolor{lightgray}
$\varphi_6$ (1/1)
& 0/1 & 0/1 & 0/1
& 0/1 & 0/1 & 0/1
& 0/1 & 0/1 & 0/1
& 0/1 & 0/1 & 0/1
& 0/1 & 0/1 & 0/1
& 0/1 & 0/1 & 0/1
& 0/1 & 0/1 & 0/1
& 0/1 & 0/1 & 0/1 \\

$\varphi_7$ (1/1)
& \textbf{1/1} & \textbf{1/1} & \textbf{1/1}
& \textbf{1/1} & \textbf{1/1} & \textbf{1/1}
& \textbf{1/1} & \textbf{1/1} & \textbf{1/1}
& \textbf{1/1} & \textbf{1/1} & \textbf{1/1}
& \textbf{1/1} & \textbf{1/1} & \textbf{1/1}
& \textbf{1/1} & \textbf{1/1} & \textbf{1/1}
& \textbf{1/1} & \textbf{1/1} & \textbf{1/1}
& \textbf{1/1} & \textbf{1/1} & \textbf{1/1} \\

\midrule

\textbf{Total} (15/21)
& 13\% & 15\% & 22\%
& 33\% & 38\% & 44\%
& 13\% & 7\% & 22\%
& \textbf{80\%} & \textbf{84\%} & \textbf{77\%}
& 20\% & 23\% & 22\%
& 33\% & 38\% & 44\%
& 6\% & 7\% & 11\%
& \textbf{80\%} & \textbf{84\%} & \textbf{77\%} \\

\bottomrule
\end{tabular}
\end{adjustbox}

\caption{Coverage comparison across  coverage types (cov$_1$--cov$_3$) for each method and agent under identical  settings.}
\label{tab:coverage_breakdown}
\end{table*}

\Cref{tab:coverage_breakdown} presents the coverage of the specifications across all treatments.
Overall, under $cov_1$ \stada achieves the highest coverage at 12/15 (80\%) for both agents, outperforming both versions of \carlaTenX (5/15; 33\%) by 47 percentage points (pp), \carlabase (transfuser) (2/15; 13\%) by 67 pp, \carlabase (interfuser) (3/15; 20\%) by 60 pp, ScenicNL (transfuser) (2/15; 13\%) by 67 pp and ScenicNL (interfuser) (1/15; 6\%) by 74 pp. 
Similarly, \stada significantly outperformed all other treatments under both $cov_2$ and $cov_3$. 
This strong performance demonstrates \stada's ability to automatically generate tests that cover the preconditions of a complex set of specifications.  
We discuss the test coverage of selected specifications below in more detail.

% \matt{Instead of going into each one by one can you group them based on similarities and then discuss the groupings (3 and 4, perhaps group 1 and 6 since no technique does well on those.  This would save space and you will have to compress a bit.} \joy{3, 4 are grouped. I grouped 5, 6 because the specifications are very similar in nature. The discussion for 1 is different to 5, 6. I can merge 1 with 5, 6 if we need space later.}

For $\varphi_0$, ScenicNL repeatedly generated arbitrary intersections rather than stop sign-controlled ones. Since most of the CARLA intersections are controlled by traffic signals, the likelihood of encountering stop signs was inherently low, reducing ScenicNL’s probability of satisfying $\varphi_0$.
On the other hand, \stada generated intersections with stop signs in every simulation of $\varphi_0$, illustrating \stada's fine-grained control over object attributes, which is particularly impactful when the required attributes are rare. 

$\varphi_1$ is time-sensitive, meeting the precondition 
%depends on the timing of the driving agent.
%It states that \textit{Ego should yield the right-of-way to the vehicle on its right if both arrive at approximately the same time}.
%More specifically, it 
requires that the 
% ego
vehicles
arrive at the intersection simultaneously.
%, i.e., at the exact same frame with another vehicle. 
%Even though \stada positions both vehicles close to an intersection.
%, it depends on the AV driving agent and autopilot to enforce the vehicles reaching the intersection together. 
Since entering the intersection on the exact same frame is extremely unlikely, none of the treatments cover it within the simulated budget. 
To relax the simultaneity requirement, we replaced the $\mathbf{X}$ in $\varphi_1$ with $\mathbf{F}$ and created $\varphi_1'$ to specify that, eventually, instead of the next frame, both vehicles were at the intersection. Formally, at frame \textit{n}, both vehicles are away from the intersection, and at frame \textit{n+h}, both vehicles are at the intersection with the referenced vehicle positioned on the right of the ego.  \stada (transfuser) and \stada (interfuser) cover $\varphi_1'$ with minimum values of $h=3$ and $h=8$ frames respectively, which correspond to 0.15 and 0.40 seconds at 20 Hz. No other treatment could cover $\varphi_1'$.

For $\varphi_3$ and $\varphi_4$, negating the conjunctive
constraint that defines \textit{following another vehicle too closely} results in 7  ways to satisfy the specifications, 4 of which are physically realizable.
\stada covered (4/4; 100\%) for $\varphi_3$ and 
(3/4; 75\%) for $\varphi_4$, whereas the best baseline only covered 2/4.

$\varphi_5$ and $\varphi_6$ are the most challenging preconditions to cover because they require 
% for
the
ego to overtake another vehicle in front of it. Only \stada could cover $\varphi_5$, and no technique could cover $\varphi_6$. For $\varphi_5$, it is sufficient that the ego overtakes a bike on the same lane. However, driving agents like Transfuser and Interfuser typically maintain vehicle-following behavior rather than proactive overtaking. For $\varphi_6$, even though the \ltlf does not mention it explicitly, it requires the ego to overtake another vehicle using the opposing lane, which the SUTs were not trained to do.
%Transfuser  was trained using a rule based \textit{expert agent} which itself is designed to avoid overtaking using the opposing lane when blocked by another vehicle. 
%Moreover, prior benchmarking studies \cite{li2025comprehensiveevaluationendtoendai} show that popular benchmark datasets like Longest6, NEAT, Langauto, etc do not include overtaking scenarios. 
Thus, even though \stada generates paths over the opposing lane, the agents prioritize safety over path adherence and refrain from overtaking. 

Notably, the coverage for \stada and \carlaTenX do not vary based on the SUT, while
% Notably, for both SUTs, \stada and \carlaTenX achieve identical coverage, while
\carlabase and ScenicNL show comparable performance across agents. 
This is caused by the fact that both SUTs exhibit a similar capability of following trajectory guidance.

\begin{tcolorbox}[colback=gray!10, colframe=black!50, boxrule=0.5pt, arc=2pt]
\textbf{Answer to RQ1.}
\stada is highly effective, producing simulations that cover the \ltlf scenario space, achieving improvements of 47–74 pp in $cov_1$, 46–77 pp in $cov_2$, and 33-66 pp in $cov_3$ over the baselines.
\end{tcolorbox} 

\subsection{\textit{RQ2: Efficiency}}

\begin{figure}[t]
    \centering
    \includegraphics[width=\columnwidth]{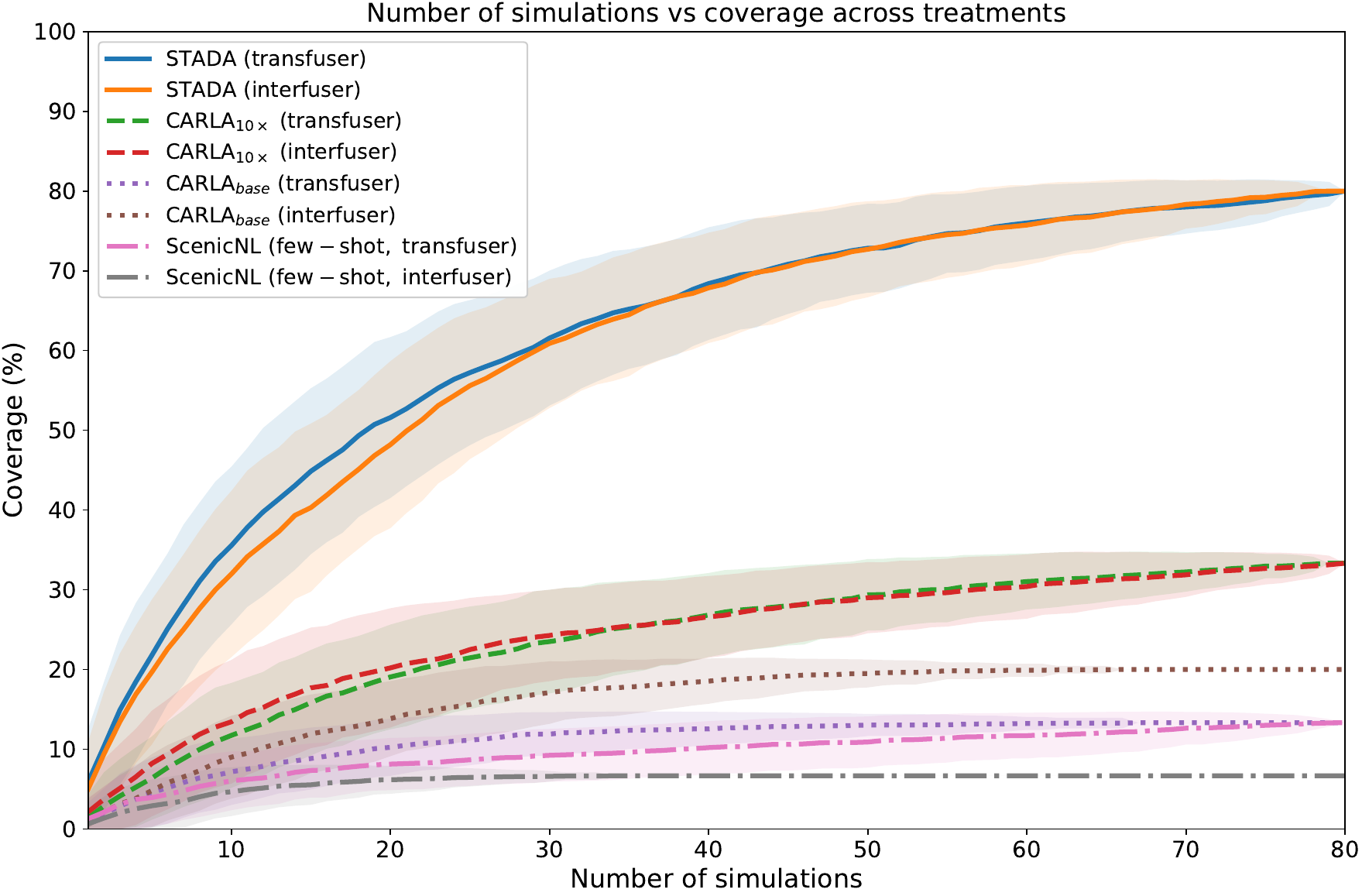}
    \caption{Mean coverage ($cov_1$) vs. number of simulations across different treatments.}
    \label{fig:coverage}
\end{figure}

\Cref{fig:coverage} shows $cov_1$ as a function of simulation count across different treatments for Transfuser and Interfuser. For each treatment, we sample 200 random permutations of the 80 simulations and compute the mean (lines) and standard deviation (shaded areas) of coverage at each step. 
%The solid curve represents the mean, while the shaded region denotes the variability around the mean (± one standard deviation).
%Here, we assess the efficiency of all of the treatments in terms of the number of simulations to reach certain coverage levels. 
\stada rises the fastest and dominates other treatments. Within about 10-12 simulations, \stada passed the second best method \carlaTenX's maximum coverage of 33\%.  \stada shows a steep initial gain of 50\%–55\% in the first 20–25 simulations and steady gains afterwards, demonstrating 
%that coverage has not saturated and showing the utility of \stada to continue generating 
%useful simulations that can 
potential to cover the missing scenarios like the one required for $\varphi_4$. 
% \matt{We need to think about what we want to highlight.  First we need to end this section with an answer to the RQ.  I don't think that belaboring how bad the other models were is worthwhile - since you already showed that they achieve low coverage in RQ1} \joy{added RQ2 answer, reduced belaboring how bad the other models are. The following paragraph can also be removed.}
The coverage achieved by \carlaTenX, despite having 10 times the entity resources, remains well below \stada, indicating that brute force scaling alone is not sufficient and also sometimes can negatively impact the coverage, e.g., a road that is too crowded can limit lane change possibilities.
%and encourage conservative straight driving behavior with minimal variations. 

% \matt{This para does say something interesting.  It is important I think to make comments about whether the SUT has any affect and this is the first time you've done that.}
%Importantly, the relative ranking of the treatments remains consistent across driving agents, indicating that the performance gain primarily arises from the test generation approach rather than the specific driving model.

\begin{tcolorbox}[colback=gray!10, colframe=black!50, boxrule=0.5pt, arc=2pt]
\textbf{Answer to RQ2.}
\stada is efficient, attaining higher coverage with fewer simulations than competing methods showing plateauing trends.
\end{tcolorbox}

\section{Conclusion}
\label{sec:conclusion}
This work addresses the question of how to thoroughly test control agents relative to a formal requirement in a realistic simulation environment.
\stada achieves this by decomposing \ltlf requirements into a set of graphs that partition its behavior.
For each behavior, it generates constraints on the initial scenes and trajectories through the space defined by those scenes that are consistent with the behavior.  From this information, it generates simulation code that can be used to instantiate targeted simulations aiming to cover the behavior of the requirements within which agent behavior can be monitored. 
A broad evaluation across 8 specifications, 3 coverage criteria, and 2 agents under test demonstrates that \stada significantly outperforms
3 state-of-the-art baselines.
This provides strong evidence that specification-based test generation for complex temporal requirements can add value to validation processes for autonomous driving agents.

%ScenicNL, powered by GPT-4, and \carlaTenX, which deploys an order of magnitude more vehicles, both failed to match \stada’s coverage performance, highlighting the limitations of LLM-based simulation code generation and brute-force scenario exploration. These findings suggest that, despite the impressive capabilities of LLMs in general code generation tasks, producing domain-specific simulation scripts remains challenging, and simply scaling traffic density is not an effective substitute for systematic, specification-driven test generation.

%All of our experiments are executed using \scenic and the Carla simulator with Transfuser and Interfuser agents. Although these are the most widely used tools in this domain, this setup poses a threat to external validity, as the findings may vary in other tools, simulation frameworks, and agents.

%For future work, we aim to investigate the direction of dynamic adjustment of NPC behavior, where the NPC will actively change its speed and trajectory in real-time to sync with the ego. Additionally, test generation for multiple ego controlled by multiple agents simultaneously can be promising.

% \section{Acknowledgments}
% This research is supported in part by NSF awards 2312487 and 2403060; by ARO grant W911NF-24-1-0089; and by Lockheed Martin Advanced Technology Labs. The authors acknowledge Research Computing at The University of Virginia for providing access to computational resources.

\bibliographystyle{IEEEtran}
\bibliography{IEEEabrv,mybibfile}

\end{document}